# Achieving Wave Pipelining in Spin Wave Technology

Abdulqader Mahmoud[1], Frederic Vanderveken[2], Christoph Adelmann[2], Florin Ciubotaru[2], Said Hamdioui[1], Sorin Cotofana[1]
[1]Computer Engineering Laboratory, Delft University of Technology, 2628 CD Delft, The Netherlands
[2] IMEC, 3001 Leuven, Belgium
[1]E-mail: a.n.n.mahmoud@tudelft.nl

**Abstract**

By their very nature, voltage/current excited Spin Waves (SWs) propagate through waveguides without consuming noticeable power. If SW excitation is performed by the continuous application of voltages/currents to the input, which is usually the case, the overall energy consumption is determined by the transducer power and the circuit critical path delay, which leads to high energy consumption because of SWs slowness. However, if transducers are operated in pulses the energy becomes circuit delay independent and it is mainly determined by the transducer power and delay, thus pulse operation should be targeted. In this paper, we utilize a 3-input Majority gate (MAJ) to investigate the Continuous Mode Operation (CMO), and Pulse Mode Operation (PMO). Moreover, we validate CMO and PMO 3-input Majority gate by means of micromagnetic simulations. Furthermore, we evaluate and compare the CMO and PMO Majority gate implementations in term of energy. The results indicate that PMO diminishes MAJ gate energy consumption by a factor of 18. In addition, we describe how PMO can open the road towards the utilization of the Wave Pipelining (WP) concept in SW circuits. We validate the WP concept by means of micromagnetic simulations and we evaluate its implications in term of throughput. Our evaluation indicates that for a circuit formed by four cascaded MAJ gates WP increases the throughput by 3.6x.

## 1. Introduction

For the last 20 years, the information technology revolution has led to exploding market needs for efficient data processing [1]. CMOS technology downscaling has been enough to meet these requirements, however, due to multiple hurdles [2] it became difficult, and thus it is predicted that Moore's law will soon come to an end. Therefore, different technologies have been investigated (e.g., graphene [3], spintronics [4]) and among them Spin Wave (SW) stands apart as one of the most promising avenues because of its: i) ultra-low energy consumption, ii) acceptable delay, and iii) high scalability. As a result, it is of great interest to design circuits using SW, and to study the applicability of computing techniques, e.g., pipelining.

As such, in view of its great potential, numerous SW based logic gates and circuits have been recently reported [5]–[13]. Single output logic gates such as (N)AND, (N)OR, and X(N)OR were introduced in [5]–[7], whereas multi-output logic gates were suggested in [8]–[10]. On the other hand, different circuits, e.g., upper/lower threshold and minimum/maximum operators, were discussed at the conceptual level without validation in [11], 2-bit inputs SW multiplier was designed and validated by means of micromagnetic simulations in [12]. Last but not least, the wave pipelining of SW Majority gate-based circuits was discussed but not validated in [13]. However, despite those advances, SW technology is still at an early stage of development and more complex designs should still be investigated.

This paper proposes, validates, and evaluates a SW 3-input Majority gate under both Continuous and Pulse Mode Operation. In addition, it utilizes Pulse Mode Operation (PMO) to introduce, validate, and evaluate Wave Pipelining into SW circuits. Its main contributions can be summarized as follows:

- SW Majority (MAJ) gate design and verification: a 3-input triangle shaped SW MAJ is designed to investigate its behavior under PMO and Continuous Mode Operation (CMO).
- MAJ functionality validation and performance evaluation under CMO and PMO: MuMax3 software is utilized to validate the MAJ gate. In addition, MAJ performance is assessed under CMO and PMO, which indicates that PMO diminishes the gate energy consumption by a factor of 18.
- SW Wave Pipelining (WP) logic gate development and verification: WP concept is introduced by making use of a MAJ gate under PMO. Accordingly, a circuit formed by four cascaded MAJ gates are instantiated to examine WP's validity in SW circuits.
- WP concept validation and performance evaluation: MuMax3 software is used to validate the correct functionality of the SW WP concept. Furthermore, the due to WP throughput enhancement is evaluated, which indicates a 3.6x improvement for the considered circuit.

The rest of the paper is organized as follows. Section 2 briefly presents SW basic concepts and computing paradigm. Section 3 illustrates the SW Majority gate operation principle under CMO and PMO. Section 4 discusses WP achievement in SW circuits and evaluate its throughput impact. Section 5 explains variability and thermal noise possible effects and Section 6 concludes the paper.

## 2. Spin Wave Technology Basics

Spin wave fundamentals and computing paradigm are explained in the following lines.

When the magnetic material magnetization is out of equilibrium, its dynamics caused by the magnetic torque can be described using Equation (1), which is called the Landau-Lifshitz-Gilbert (LLG) equation [14].

$$\frac{d\vec{M}}{dt} = -|\gamma|\mu_0(\vec{M} \times \vec{H}_{eff}) + \frac{\alpha}{M_s}\left(\vec{M} \times \frac{d\vec{M}}{dt}\right) \quad (1)$$

Where $\gamma$ is the gyromagnetic ratio, $\mu_0$ the magnetic constant, M the magnetization, $H_{eff}$ the effective field, $\alpha$ the damping factor, and $M_s$ the magnetic saturation. The effective field consists of different magnetic interactions and it represents in this work the summation of the demagnetizing field, the exchange field, the external field, and the magneto-crystalline field.

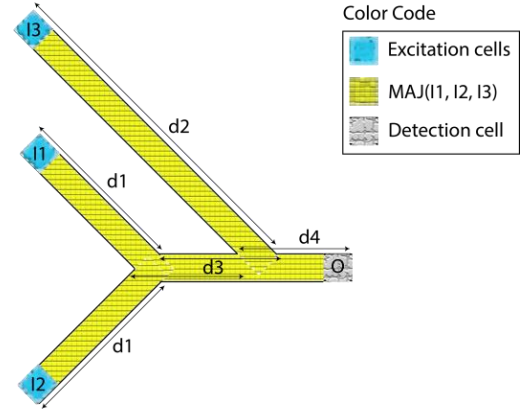

**Figure 1:** 3-input SW Majority Gate.

If the electron spins are deflected from their positions by one of the excitation methods (specified at the end of this Section), LLG equation can be linearized and results in wave-like solutions. These solutions are known as Spin Waves, which can be defined as the collective excitation of the spins in the magnetic material, and it is described by SW amplitude, phase, frequency, wavelength, and wavenumber as any other wave. The relation between frequency and wavenumber, i.e., the so-called dispersion relation, is very important for SW circuit design as it determines the SW excitation frequency, wavenumber, and wavelength [15].

The information can be encoded in the amplitude, and phase at multiple frequencies in the SW technology [10], [14]. The interaction between SWs is governed by the interference principle. For instance, if SWs with the same amplitude, frequency, and wavelength meet at an intersection point, they interfere constructively if they have the same phase $\Delta\phi = 0$, whereas if they are out of phase, i.e., $\Delta\phi=\pi$, they interfere destructively [10], [14]. In addition, SW interference provides natural support for Majority function evaluation. For example, if 3 SWs with the same amplitude, frequency, and wavelength propagate through waveguides, the interference result is based on a Majority decision, i.e., the resulting SW has phase $\Delta\phi=0$ if no more than one input SW has a phase $\Delta\phi=\pi$, and a phase $\Delta\phi=\pi$ if at most one input SW has a phase $\Delta\phi=0$. Thus, one single SW device evaluates 3-input Majority [10], [14], while a CMOS counterpart needs 18 transistors. Note that logic 1 represents $\Delta\phi=\pi$, and $\Delta\phi=0$ reflects logic 0 in this paper [10], [14].

In general, a SW device consists of four regions [14]:

- Excitation: SWs are excited at this stage by means of voltage/current-driven transducers, e.g., magnetoelectric cells, microstrip antennas, spin orbit torques [14].
- Waveguide: The medium where SWs propagate. Different magnetic materials can be used to build it, e.g., Permalloy, Yttrium iron garnet, CoFeB. Depending on the used materials, the SW properties vary such as SW group velocity and dispersion relation [14].
- Functional Region: It is the section where SWs are manipulated, i.e., interfere with each other and/or are amplified or normalized [14].
- Detection: The output SW is detected at this stage by means of voltage/current-driven transducers, e.g., magnetoelectric cells, microstrip antennas, spin orbit torques [14].

## 3. SW Majority Gate Operation Mode

In the following lines, the 3-input Majority gate working under Continuous Mode Operation (CMO) and Pulse Mode Operation (PMO) is explained. Also, the simulation parameters, in simulation results, and performance evaluation are discussed.

### 3.1. CMO and PMO Spin Wave MAJ Concept

Figure 1 presents the 3-input SW Majority gate we utilize as discussion vehicle to demonstrate and evaluate the impact of the two operation modes. Note that this gate is a simplified version of the one in [16], and that both modes can be also demonstrated on different structures like [8], [17].

In order to achieve the desired functionality, the Majority gate parameters and dimensions must be carefully determined. SW wavelength $\lambda$ must be larger than the waveguide width in order to simplify the interference pattern. In addition, to correctly capture the output values the input SWs must be excited at the same time and with the same amplitude, frequency, and wavelength. Further, the structure dimensions $d_1$, $d_2$, $d_3$ are essential for the gate behavior. For instance, if the SWs should constructively interfere when they are in-phase, and destructively when they are out-of-phase, $d_1$, $d_2$, $d_3$ must be $n\lambda$ (where n = 0, 1, 2, 3, …). In contrast, if the desired case is to interfere destructively when they are in-phase, and constructively when they are out-of-phase, $d_1$, $d_2$, $d_3$ must be $(n+1/2)\lambda$ (where n = 0, 1, 2, 3, …).

In order to detect the output correctly, $d_4$ must be also accurately chosen as its value determines if the gate computes the Majority function or its complement. For example, if the gate output itself is desired, $d_4$ must be $n\lambda$ and if inverted Majority is the desired output, $d_4$ must be $(n+1/2)\lambda$.

As seen from the point of view of SW excitation SW gates and circuits can operate in two main modes: Continuous Mode Operation (CMO) and Pulse Mode Operation (PMO).

- CMO: SWs are excited by using continuous signals such that the excitation signal remains active until the output result detection completion because: i) SW with fixed frequency and wavelength are excited in the same time with the same amplitude, and ii) The SW resulted from input SWs interferences can be easily detected as it has the same frequency and wavelength as the inputs. However, as the excitation signal is active from the beginning to the end of the calculation, CMO SW devices energy consumption is large, especially because they are slow as further discussed in Subsection 3.4.

- PMO: In contrast to CMO, this operation is more energy effective as a result of the fact that the excitation signal is active only for a short period of time. However, PMO is more complex to work with because: i) SW excitation by means of a pulse signal produces multiple SWs with different wavelengths and frequencies as the pulse covers a large band in the frequency domain. Therefore, all the covered frequencies are excited in the waveguide, which results in the creation of multiple wavenumbers ($k = 2 \times \pi/\lambda$) and wavelengths SWs.

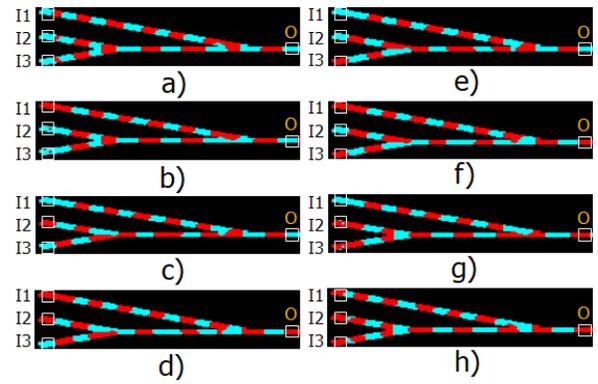

**Figure 2:** CMO 3-input SW MAJ MuMax3 Simulation.

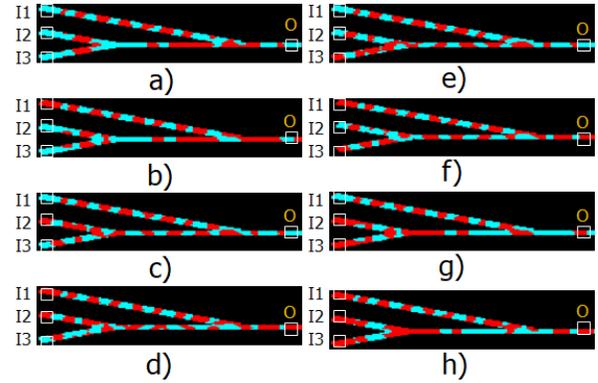

**Figure 3:** PMO 3-input SW MAJ MuMax3 Simulation.

However, the number of frequencies can be limited by using sinusoidal or Gaussian excitation signals. ii) If the inputs are not located at the same distance from the output, the input SWs must be excited at different time moments and with different amplitudes. Otherwise, the gate malfunctions as the closest input SW reaches the output before the furthest SW, which also has a lower energy when reaching the output as a result of the damping effect in the waveguide. This requires a complex clocking scheme, which can be simplified by attempting to equalize all the input to output propagation paths by means of proper layout and/or delay buffers.

The generation of the MAJ gate output O value (see Figure 1) is performed as follows: SWs are excited at $I_1$, $I_2$, and $I_3$ to propagate through the waveguides using continuous signal which is active until detecting the result (CMO) or pulse signals (PMO). The SW resulted from the interferences between input SWs are detected based on phase such that if the output SW has a phase difference of 0 versus a predefined phase, the output is logic 0, whereas the output is logic 1 if the phase difference is $\pi$.

### 3.2. Simulation Setup

We use MuMax3 [18] to validate the correct functionality of the proposed concepts and structures. The MAJ gate is built with 50 nm wide $Fe_{60}Co_{20}B_{20}$ waveguide along with the following parameters [19]: Magnetic saturation $M_s$=1100 kA/m, Perpendicular anisotropy constant $k_{ani}$=830 kJ/m$^3$, damping constant α=0.004, Exchange stiffness $A_{exch}$=18.5 pJ/m, and Thickness t=1 nm. For the CMO we choose a SW wavelength of 55 nm as for proper gate operation it has to be larger than the waveguide width w. Accordingly, Figure 1 structure dimensions were determined as: $d_1$=330 nm (n = 6), $d_2$=880 nm (n = 16), $d_3$=220 nm (n = 4), and $d_4$ = 55 nm (n = 1). Furthermore, using the parameters in Table I and w, the we calculate the SW dispersion relation [15] and determine SW frequency f = 10 GHz and wavenumber k=2π/λ=50 rad/μm. As the target is to compare the two operation modes, we maintain the same dimensions also for PMO.

### 3.3. Simulation Results

Figure 2 presents the simulation results of the 3-input Majority gate (Figure 1) working under CMO for $\{I_1,I_2,I_3\}$= {0,0,0}, {0,0,1}, {0,1,0}, {0,1,1}, {1,0,0}, {1,0,1}, {1,1,0}, and {1,1,1}, respectively, from a) to h). Note that in the Figure logic 0 and 1 are represented by blue and red color, respectively. As mentioned previously, the input SWs activation signal is on all the time and keeps exciting SW until the output O detection is completed. As it can be observed from Figure 2, the output O is correctly detected. For example, O is logic 0 for the input patterns $\{I_1,I_2,I_3\}$= {0,0,0}, {0,0,1}, {0,1,0}, and {1,0,0}, whereas O = 1 for the input combinations $\{I_1,I_2,I_3\}$= {0,1,1}, {1,0,1}, {1,1,0}, and {1,1,1}, which indicate that the Majority gate behaves correctly under the CMO scenario.

Figure 3 presents the simulation results of the 3-input Majority gate (Figure 1) working under PMO for $\{I_1,I_2,I_3\}=$ $\{0,0,0\}$, $\{0,0,1\}$, $\{0,1,0\}$, $\{0,1,1\}$, $\{1,0,0\}$, $\{1,0,1\}$, $\{1,1,0\}$, and $\{1,1,1\}$, respectively, from a) to h). In this case, we make use of a 100 ps sinusoidal excitation signal in order to decrease the number of excited frequencies. As it can be observed from Figure 3, O is still correctly detected. For example, the O = 0 for the input patterns $\{I_1,I_2,I_3\}=$ $\{0,0,0\}$, $\{0,0,1\}$, $\{0,1,0\}$, and $\{1,0,0\}$, whereas O = 1 for the input combinations $\{I_1,I_2,I_3\}=\{0,1,1\}$, $\{1,0,1\}$, $\{1,1,0\}$, and $\{1,1,1\}$, which indicate which indicate that the Majority gate behaves correctly under the PMO scenario. As it can be observed in the Figure, PMO generates multiple SW wavelengths as the SWs do not travel the same distance and because of that and in contrast to CMO, results are not accurate in all positions. However, as the output position is accurately determined, results are correctly captured at the output position as depicted in the Figure. To conclude, the simulation results demonstrate that the Majority gate works correctly under both CMO and PMO scenarios.

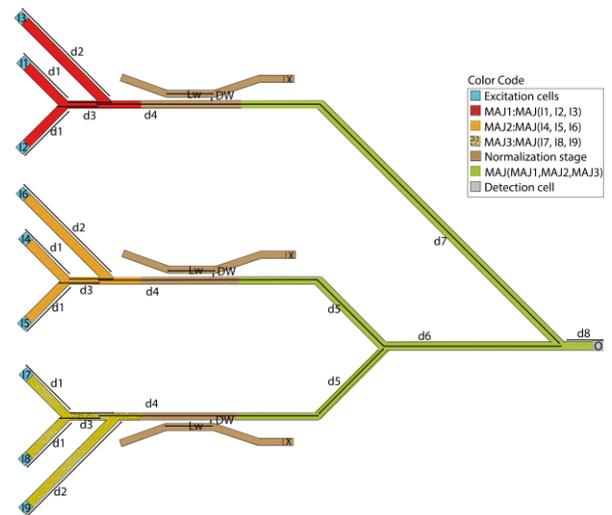

**Figure 4:** 4 Cascaded 3-input SW MAJ Gates.

### 3.4. Performance Evaluation

To get inside on the practical implications of operation mode on the MAJ gate we its energy consumptions under both CMO and PMO. To this end the following assumptions are in place [13]: the single SW transducer exhibits a 0.42 ns delay and a power consumption of 34.3 nW. Note that because we analyze one gate only, we do not include clock complexity and overhead in our evaluation. However, clock will certainly play an important role when analyzing large complex circuits with un-equal paths. Moreover, as SW technology is immature, the made assumptions might not hold true as it evolves towards maturity and as such re-evaluation might be of interest.

Note that the SW propagation delay is extracted directly from MuMax3 simulations, and it is 1 ns. Thus, by adding the input and output transducers delays to the SW propagation delay the SW MAJ gate total delay sums up to 1.84 ns.

Under these assumptions, as the source is active for 100 ps, and there are four transducers, PMO 3-input SW MAJ induces a 13.7 aJ energy consumption. CMO 3-input SW MAJ results in an energy consumption of 252.5 aJ as the source is active for 1.84 ns and there are four transducers. Hence, PMO diminishes the energy consumption by a factor of 18x. This can be easily explained by the fact that when SWs excitation is performed by the continuous application of voltages/currents the overall energy consumption is determined by the transducer power and the gate critical path length (delay). However, if transducers are operated in pulses the energy becomes gate delay independent as it is mainly determined by the transducer power and delay, thus pulse operation should be targeted. Note that regardless of the operation mode SWs are excited and detected at different time moments, which makes clocking unavoidable. However, its complexity analysis requires further SW technology developments and constitutes future work.

While PMO has a substantial impact on energy consumption reduction it is also an enabling factor for the realization of SW circuits operating under the Wave Pipelining paradigm [20][21], which increases circuit throughput without requiring inter-stage registers. Based on this observation, the next section introduces, validates, and evaluates SW Wave Pipelined circuits.

## 4. Wave Pipelining Achievement in SW Technology

This section introduces the Wave Pipelining (WP) concept in the context of SW technology and discuss it on a simple circuit composed by 3 cascaded MAJ gates. In addition, micro-magnetics simulation results and performance evaluation are also presented.

### 4.1. Wave Pipelining

Pipelining is processing multiple sets of inputs before the first set reaches the output [20]. This is performed by slicing the function into multiple stages where each stage is isolated from the previous and the next stage by means of registers that store intermediate processed data [20]. Each stage processes a data set independently from the other stages. After data sets processing is complete, the results are stored in a register to be passed to the next stage on the following clock cycle [20]. In order to minimize (preferably totally remove) the register number in a pipelined system, Wave Pipelining (WP) was introduced [21]. The main idea is to allow for the coexistence and interference free handling of multiple data sets withing a register free processing pipeline circuit [21]. To be able to operate in such a manner the circuit has to be redesigned such that all its propagation paths exhibit the same delay. This guaranties that input sets do not interfere within the circuit and reach the output in their chronological order.

## 4.2. Wave Pipelined SW Circuits

Figure 4 presents the SW circuit we make use of as WP discussion vehicle. It comprises 4 MAJ gates and 3 directional couplers [12], has 9 inputs $I_1, I_2, I_3, I_4, I_5, I_6, I_7, I_8, I_9$, and one output O, which evaluates MAJ(MAJ($I_1, I_2, I_3$), MAJ($I_4, I_5, I_6$), MAJ($I_7, I_8, I_9$)). Note that to allow for input SWs excitation at the same moment in time and with the same amplitude, all inputs have to be places at the same distance from the gate output. Moreover, as the MAJ gate SW output has an input data depended amplitude and a cannot be directly cascaded, we make use of 3 directional couplers to normalize the output of the layer one MAJ gates [12].

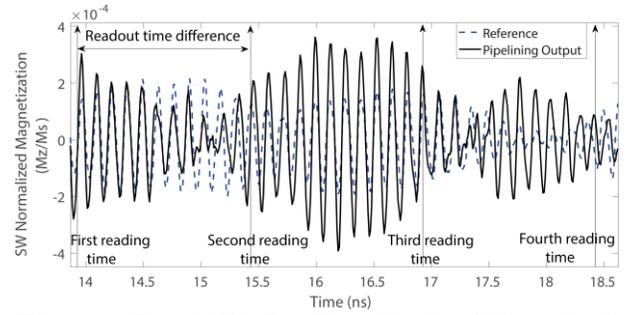

**Figure 5:** First 4 SWs Sets Wave Pipelined Normalized Magnetization.

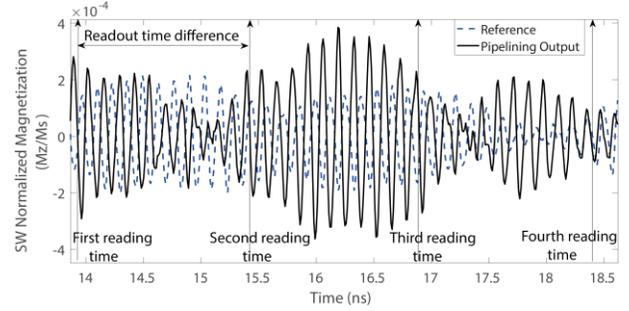

**Figure 6:** Second 4 SWs Sets Wave Pipelined Normalized Magnetization.

The aforementioned in Subsection 3.1 design steps hold true for this case as well. In addition, the directional couplers must be designed, i.e., determine the coupler length Lw and the gap between the directional coupler and the main waveguide DW, such that they normalize the output of the layer one MAJ gates and the layer two can properly operate. This can be performed by making use of the equations in [12].

To detect the output O (see Figure 4) correctly the operation principle of the proposed circuit is as follows: SWs are excited at each source $I_1, I_2, I_3, I_4, I_5, I_6, I_7, I_8$, and $I_9$. Then the SWs interfere in groups of three per MAJ gate and the resulted SWs are normalized using the directional couplers. After that, the three SWs produced by the directional couplers interfere constructively or destructively depending on their phases and, finally, the resulted SW is detected at the output by means of phase detection.

As we want to utilize the structure in Figure 4 to demonstrate the SW WP concept, the PMO must be utilized as it is the WP enabling factor because a new input set can be applied before the evaluation of the previous one is completed. Therefore, the sources $I_1, I_2, I_3, I_4, I_5, I_6, I_7, I_8$, and $I_9$ are utilized to excite multiple SWs' sets by means of pulse signals distanced by a certain time gap determined such that each set doesn't affect the previous excited set. In this way multiple data sets SWs coexist in the circuit, properly interfere in the level 1 MAJ gates, are normalized, interfere again in the level 2 MAJ gate, and the corresponding output value is detected at O. As such the circuit throughput is increased by a time gap determined factor without interstage register insertion.

## 4.3. Simulation Setup

To validate the correct functionality of the circuit in Figure 4 under WP operation we make use of MuMax3 simulations while keeping the waveguide material and width, wavelength and parameters reported in Subsection 3.2. Consequently, Figure 4 structure dimensions are: $d_1$=330 nm (n = 6), $d_2$=880 nm (n = 16), $d_3$=220 nm (n = 4), $d_4$=2750 nm (n = 50), $d_5$=935 nm (n = 17), $d_6$=3300 nm (n = 60), $d_7$=2145 nm (n = 39), and $d_8$=55 nm (n = 1). The directional coupler dimensions are determined based on the equations in [12] as Lw=1500 nm and DW =10 nm.

## 4.4. Simulation Results

Theoretically speaking, as the input transducers operate under 0.5 ns pulses the maximum time gap between two consecutive data sets is also 0.5 ns. However, based on MuMax3 simulation, this is not achievable with the utilized waveguide material, parameters, pulse duration, and pulses strength that allow the SWs to persist in the waveguide for more than 5 ns, which means that an already excited SW might affect new coming SWs. Based on experiments we identified 1.5 ns as the optimal time distance between two consecutive sets for this material, parameters, pulse duration, and pulses strength. Thus, we evaluated the circuit in Figure 4 by means of input data sets applied at 1.5 ns time gap instead of 0.5 ns. As the circuit has 9 inputs there are 512 possible input combinations, but due to symmetry many of them result in the same output. As such, we have chosen 8 input combinations that fully exercise the second layer gate, i.e., $\{I_1,I_2,I_3,I_4,I_5,I_6,I_7,I_8,I_9\}$={0,0,0,0,0,0,0,0,0},{0,0,0,0,0,0,1,1,1},{0,0,0,1,1,1,0,0,0}, {0,0,0,1,1,1,1,1,1}, {1,1,1,0,0,0,0,0,0}, {1,1,1,0,0,0,1,1,1},{1,1,1,1,1,1,0,0,0}, and{1,1,1,1,1,1,1,1,1 }, which induce {II1,II2,II3} = {0,0,0}, {0,0,1}, {0,1,0}, {0,1,1}, {1,0,0}, {1,0,1}, {1,1,0}, and {1,1,1}, where II1, II2, and II3 are the normalized outputs of the first, second, and third layer 1 MAJ gate, respectively. For the same material related reasons, we split the 8 combinations into two groups and apply them separately to the circuit. In this way each new set in the group is still slightly affected by the previous set, but the circuit still functions correctly if up to four sets are injected by using 0.5 ns pulse signal and a time gap of 1.5 ns.

Based on this timing scheme SWs sets are excited starting from 0 ns to 0.5 ns, from 1.5 ns to 2.0 ns, from 3.0 ns to 3.5 ns, and from 4.5 ns to 5.0 ns and the corresponding MuMax3 simulation results are presented in Figure 5. In order to validate the proper WP behavior, we make use of a reference SW, which is the result of exciting the input SWs with {0,0,0,0,0,0,0,0,0} using a pulse signal with a duration of 0.5 ns. As it can be observed from the Figure, the output SW has a phase difference of 0 at time equals to 13.9 ns. As the SW sets are pipelined with 1.5 ns time difference, the result for the next set should be ready after 1.5 ns. Therefore, the next set result is detected at time equals to 15.4 ns at which the SW has a phase difference of approximately 0. Likewise, the third set result is detected after 1.5 ns, the resulted SW has a phase difference of approximately 0 at time equals to 16.9 ns. Finally, the fourth set result is detected at time equals to 18.4 ns, which has a phase difference of approximately $\pi$. Therefore, all outputs are correctly computed and detected. Similarly, the results in Figure 6 can be analyzed, but since the excited SWs are in the following order {1,1,1,1,1,1,1,1,1}, {1,1,1,1,1,1,0,0,0}, {1,1,1,0,0,0,1,1,1}, and {1,1,1,0,0,0,0,0,0}, the first three SWs sets will have a phase difference of $\pi$ with the reference signal, whereas the last one a phase difference of 0 as can be noticed in Figure 6 at times 13.9 ns, 15.4 ns, 16.9 ns, and 18.4 ns, respectively. Note that the same reference signal in Figure 5 is utilized.

To conclude, the simulation results demonstrate that the WP concept is validated within the SW technology framework and that up to 4 SWs sets can be wave pipelined in the waveguide because SWs persist in the waveguide for longer than 5 ns, and therefore each new set might be affected by previous sets, but since the new excited SW is stronger, their effect is limited. Note that there might be a place for optimization for the number of the WP sets by: i) Exciting SWs sets with different energy level such that the first SWs set has the lowest energy level while the last one has the highest energy level. ii) Decreasing the pulse duration. iii) Increasing the time difference between the excited SWs sets. iv) Using a material with a slightly higher damping effect as waveguide such that SWs can still propagate on long enough distances but in the same time vanish quicker.

### 4.5. Performance Evaluation

To get inside on the SW WP potential, we examine the throughput of the circuit in Figure 4 with and without WP. In order to do so, we assume that an idle time of 5 ns is required to avoid the effect of previous input SWs on newly applied SWs. This time overhead is required after each input in normal circuit operation and after each input set group when WP is in place. From Figures 5 and 6, we can notice that the 8 cases results are ready in 36.8 ns as the group wave pipelined result is ready after 18.4 ns. However, due to the idle time the second group has to be 5 ns delayed, which implies that 8 operations can be completed in 41.8 ns. In contrast, the 8 not wave pipelined evaluations can be performed in 151.2 ns by taking into consideration that one new result is available each and every 13.9 ns and 5 ns idle time is required between successive input sets. Thus, WP utilization increases the throughput by 3.6x for the implementation on the structure in Figure 4.

### 5. Variability and Thermal Effect

In this paper, the main goal is to introduce and validate SW based Wave Pipelining without considering the thermal noise and variability effects. However, the trapezoidal waveguide cross section and edge roughness effects were presented in [22]; it was illustrated that both have limited effects, and the gate continues functioning correctly at their presence. In addition, it was presented that the thermal noise has a limited effect as well in [22] after analyzing the results at different temperatures. Therefore, it is not expected to have an impact on the gate correct functionality nor on the WP concept. However, such phenomena will be explored thoroughly in the near future.

### 6. Conclusions

We proposed and validated by means of micromagnetic simulations a SW 3-input Majority gate under continuous and pulse mode operation regimes. We also evaluated the gate energy consumption and our results indicated that Pulse Mode Operation (PMO) diminishes the gate energy consumption by a factor of 18, when compared with the Continuous Mode Operation. In addition, we also presented how PMO enables Wave Pipelining (WP) within SW circuits and validated WP on a simple circuit by means of micromagnetic simulations. Furthermore, we demonstrated that WP utilization improves the circuit throughput by 3.6x.

### Acknowledgement

This work has received funding from the European Union's Horizon 2020 research and innovation program within the FET-OPEN project CHIRON under grant agreement No. 801055. It has also been partially supported by imec's industrial affiliate program on beyond-CMOS logic. F.V. acknowledges financial support from the Research Foundation--Flanders (FWO) through grant No.~1S05719N.